\documentstyle[twocolumn,graphicx,aps]{revtex}

\begin{document}
\draft

\title{Quantum Monte Carlo study of the 3D-1D crossover for a trapped Bose gas}

\author{G.E. Astrakharchik and S. Giorgini}

\address{
{\small\it  $^{1}$Dipartimento di Fisica, Universit\`a di Trento, and Istituto 
Nazionale per la Fisica della Materia, I-38050 Povo, Italy}
\\ (\today)
\\ \medskip}\author{\small\parbox{14.2cm}{\small\hspace*{3mm}
We investigate the cross-over from three to one dimension in a Bose gas confined in highly anisotropic traps.
By using Quantum Monte-Carlo techniques, we solve the many-body Schr\"odinger equation for the ground state and 
obtain exact results for the energy per particle and the mean square radii of the cloud in the transverse and 
longitudinal direction. Results are compared with the predictions of mean-field theory obtained from the 
Gross-Pitaevskii equation and with the 1D Lieb-Liniger equation of state. We explicitly prove the occurrence 
of important beyond mean-field effects, including the appearance of Fermi-like properties as the system enters 
the Tonks gas regime. \\
\\[3pt]PACS numbers: 03.75.Fi, 05.30.Fk, 67.40.Db}} \maketitle

\narrowtext

\section{Introduction}

The study of trapped Bose systems in low dimensions is currently attracting a lot of interest. In particular, 1D
systems are expected to exhibit remarkable properties which are far from the mean-field description and are not
present in 2D and 3D. The peculiarity of 1D physics consists in the role played by fluctuations, which destroy 
long-range order even at zero temperature \cite{Schwartz}, and in the occurrence of characteristic effects due
to correlations such as the fermionization of the gas in the Tonks regime \cite{Girardeau}. 
Recent experiments with highly anisotropic, quasi-one-dimensional traps have shown first evidences of 1D features 
in the aspect ratio and energy of the released cloud \cite{Mit,Paris} as well as in the coherence properties of 
condensates with fluctuating phase \cite{Konstanz}. From a theoretical viewpoint, the emergence of 1D effects in 
the properties of binary atomic collisions, by increasing the confinement in the transverse direction, has been 
pointed out in \cite{Olshanii}. In the case of harmonically trapped gases, the occurrence of various regimes 
possessing true or quasi-condensate and the possibility of entering the Tonks gas regime of impenetrable bosons 
has been discussed in \cite{Petrov}. 

The ground-state properties and excitation spectrum of a homogeneous 1D system of bosons interacting through a 
repulsive contact potential have been calculated exactly by Lieb and Liniger long time ago \cite{Lieb}. For a 
fixed interaction strength the Lieb-Liniger equation of state reproduces in the high density regime the mean-field 
result obtained using the Bogoliubov model and in the opposite limit of low density coincides with the ground-state 
of impenetrable bosons \cite{Girardeau}. For 1D systems in harmonic traps, the exact many-body ground-state 
wavefunction in the Tonks regime has been recently calculated \cite{Girardeau2}, and the equation of state 
interpolating between the mean-field and the Tonks regime has been obtained within the local density approximation 
in \cite{Dunjko}. Methods based on local density approximation in the longitudinal direction and on the 
Gross-Pitaevskii equation for the transverse direction have been recently employed to predict the frequency of the 
collective excitations \cite{Menotti} and the ground-state energy in the 3D-1D cross-over as well as in 
the 1D mean field - Tonks gas cross-over \cite{Girardeau3}.

In this work we present exact Quantum Monte-Carlo results for the 3D-1D cross-over in harmonically trapped Bose 
gases. As a function of the anisotropy parameter of the trap we calculate the ground-state properties of the system 
and for highly anisotropic traps we point out the occurrence of important beyond mean-field effects including the 
fermionization of the gas.

\section{Theory}

We consider the following Hamiltonian 
\begin{equation}
H=- \frac{\hbar^2}{2m}\sum_{i=1}^N\nabla_i^2+\sum_{i<j}V(|{\bf r}_i-{\bf r}_j|)+\sum_{i=1}^N V_{ext}({\bf r}_i) \;,
\label{ham}
\end{equation}
describing a system of $N$ spinless bosons of mass $m$ interacting through the two-body interatomic potential $V(r)$ 
and subject to the axially symmetric harmonic external field $V_{ext}({\bf r})=m(\omega_\perp^2 r_\perp^2 + \omega_z^2 
z^2)/2$, where $z$ is the axial coordinate, $r_\perp$ is the radial transverse coordinate and $\omega_z$, 
$\omega_\perp$ are the corresponding oscillator frequencies. 

For the interatomic potential we use two different repulsive model potentials:

\noindent
1) Hard-sphere (HS) potential defined as
\begin{equation}
V(r)=\left\{  \begin{array}{cc} +\infty & (r<a)  \\
                                   0    & (r>a)  \end{array} \right. \;,
\label{HS}
\end{equation}
where the diameter $a$ of the hard sphere corresponds to the $s$-wave scattering length.

\noindent
2) Soft-sphere (SS) potential defined as    
\begin{equation}
V(r)=\left\{  \begin{array}{cc}    V_0  & (r<R)  \\
                                   0    & (r>R)  \end{array} \right. \;.
\label{SS}
\end{equation} 
The $s$-wave scattering length is given by $a=R[1-\tanh(K_0R)/K_0R]$, with $K_0^2=V_0m/\hbar^2$ and $V_0>0$. For 
finite $V_0$ one always has $R>a$, while for $V_0\to+\infty$ the SS potential coincides with the HS one with $R=a$. 
The height $V_0$ of the potential is fixed by the value of the range $R$ in units of the scattering length, for which
we choose $R=5a$. It is worth noticing that the HS and the SS model with $R=5a$ represent two extreme cases for a 
repulsive interatomic potential. In the HS case, the energy is entirely kinetic, while for the SS potential 
$a\simeq(m/\hbar^2)\int_0^{\infty}V(r)r^2 dr$, according to Born approximation, and the energy is almost all 
potential. By comparing the results of the two model potentials we can investigate to what extent the ground-state 
properties of the system depend only on the $s$-wave scattering length and not on the details of the potential.     

The relevant parameters of the problem are the number of particles $N$, the ratio $a/a_\perp$ of the scattering 
length to the transverse harmonic oscillator length $a_\perp=\sqrt{\hbar/m\omega_\perp}$ and the anisotropy 
parameter $\lambda=\omega_z/\omega_\perp$. For a given set of parameters we solve exactly, using the Diffusion 
Monte-Carlo (DMC) method, the many-body Schr\"odinger equation for the ground state and we calculate the energy per 
particle and the mean square radii of the cloud in the axial and radial directions.  Importance sampling is used 
through the trial wavefunction $\psi_T({\bf R}) \equiv \psi_T({\bf r}_1,..,{\bf r}_N) = \prod_i f_1({\bf r}_i)
\prod_{i<j}f_2(r_{ij})$, where $f_1$ and $f_2$ are respectively one and two-body Jastrow factors. For the one-body
term, which accounts for the external confinement, we use a simple gaussian ansatz $f_1(r_\perp,z)=\exp(-\alpha_\perp
r_\perp^2-\alpha_z z^2)$, with $\alpha_\perp$ and $\alpha_z$ optimized variational parameters. The two-body term
$f_2(r)$ accounts instead for the interparticle interaction and is chosen using the same technique employed in 
Ref. \cite{US} for a homogeneous system. Of course, since DMC is an exact method, the precise choice of 
$\psi_T({\bf R})$ is to a large extent unimportant and the results obtained are not biased by the choice of the 
trial wavefunction \cite{BORO}.
  
The DMC results are compared with the predictions of mean-field theory which are obtained from the stationary 
Gross-Pitaevskii (GP) equation 
\begin{equation}
\left(-\frac{\hbar^2}{2m}\nabla^2+V_{ext}({\bf r})+g(N-1)|\Phi({\bf r})|^2\right)\Phi({\bf r})=\mu\Phi({\bf r}) \;,
\label{GP}
\end{equation}
where $\Phi({\bf r})$ is the order parameter normalized to unity: $\int d{\bf r}|\Phi({\bf r})|^2=1$ and 
$g=4\pi\hbar^2a/m$ is the coupling constant. Further, finite size effects have been taken into account in the 
GP equation by the factor $N-1$ in the interaction term \cite{Esry}. In the case of anisotropic traps with $\lambda<1$, 
the GP equation (\ref{GP}) is expected to provide a correct description of the system if the transverse confinement 
is weak $a/a_\perp\ll 1$, and if the mean separation distance between particles is much smaller than the healing 
length  $1/n^{1/3}\ll\xi$, where $\xi=1/\sqrt{8\pi na}$ and $n$ is the central density of the cloud. In terms of
the linear density along $z$, $n_{1D}(z)=2\pi\int_0^\infty dr_\perp\;r_\perp n(r_\perp,z)$, this latter condition 
reads $1/n_{1D}\ll a_\perp^2/a$. If the mean separation distance between particles in the longitudinal direction 
becomes much larger than the effective 1D scattering length given by $a_\perp^2/a$ \cite{Petrov}, the mean-field 
approximation breaks down because of the lack of off diagonal long range order.   

The system enters the 1D regime when the motion in the radial direction becomes frozen. In this regime the radial 
density profile of the cloud is fixed by the harmonic oscillator ground state, resulting in a mean square radius 
which coincides with the transverse oscillator length $\sqrt{\langle r_\perp^2\rangle}=a_\perp$. Further, the energy 
per particle is dominated by the trapping potential and one has the condition $E/N-\hbar\omega_\perp\ll
\hbar\omega_\perp$. If the discretization of levels in the longitudinal direction can be neglected, i.e. if 
$E/N-\hbar\omega_\perp\gg\hbar\omega_z$, the 1D system can be described within the local density approximation (LDA). 
In this case, the chemical potential of the system is calculated through the local equilibrium equation 
\begin{equation}
\mu=\hbar\omega_\perp + \mu_{local}(n_{1D}(z))+\frac{m}{2}\omega_z^2z^2 \;,
\label{LDA}
\end{equation}   
where $\hbar\omega_\perp$ is the dominant contribution of the transverse confinement and $\mu_{local}(n_{1D})$ is the 
chemical potential corresponding to a homogeneous 1D system of density $n_{1D}$. If the ratio $a/a_{\perp}\ll 1$, the 
local chemical potential can be obtained from the Lieb-Liniger (LL) equation of state with the effective 1D coupling 
constant $g_{1D}=g/(2\pi a_{\perp}^2)$ \cite{Olshanii,Dunjko}. One finds: $\mu_{local}=\partial[n_{1D}\epsilon_{LL}
(n_{1D})]/\partial n_{1D}$, where $\epsilon_{LL}$ is the LL energy per particle. By using Eq. (\ref{LDA}) and the 
normalization condition $\int_{-\infty}^{+\infty} dz\; n_{1D}(z)=N$, one can obtain the chemical potential $\mu$ as 
a function of $N$ \cite{Dunjko}. The ground-state energy of the system with a given number of particles can then be 
calculated through direct integration of $\mu(N)$. 

If $n_{1D}a_\perp^2/a\gg 1$, the system is weakly interacting and the LL equation of state coincides with the mean-field 
prediction: $\epsilon_{LL}=g_{1D}n_{1D}/2$. In this regime, one finds the following results for the energy per particle 
\begin{equation}
\frac{E}{N}-\hbar\omega_\perp=\frac{3}{10}\left(3N\lambda\frac{a}{a_\perp}\right)^{2/3}\hbar\omega_\perp \;,
\label{1DMF1}
\end{equation}
and for the mean square radius of the cloud in the longitudinal direction
\begin{equation}
\sqrt{\langle z^2\rangle}=\left(3N\lambda\frac{a}{a_\perp}\right)^{1/3}\frac{a_\perp}{\sqrt{5}\lambda} \;.
\label{1DMF2}
\end{equation}
In the opposite limit, $n_{1D}a_\perp^2/a\ll 1$, the system enters the Tonks regime and the LL equation of 
state has the Fermi-like behavior $\epsilon_{LL}=\pi^2\hbar^2n_{1D}^2/6m$. The energy per particle and the mean
square radius of the trapped system are given in this case by the following expressions
\begin{equation}
\frac{E}{N}-\hbar\omega_\perp=\frac{N\lambda}{2}\hbar\omega_\perp \;,
\;\;\;\;\;\;
\sqrt{\langle z^2\rangle}=\sqrt{\frac{N}{2\lambda}}a_\perp \;.
\label{1DTG}
\end{equation}  
In terms of the parameters of the system, the two regimes can be identified by comparing the corresponding 
energies. The mean-field energy becomes favourable if $N\lambda a_\perp^2/a^2\gg 1$, whereas the Tonks gas 
is preferred if the condition $N\lambda a_\perp^2/a^2\ll 1$ is satisfied. 

In order to account for effects beyond local density approximation we have also applied the DMC method to a system 
of $N$ particles interacting through the Lieb-Liniger Hamiltonian in the presence of harmonic confinement
\begin{eqnarray}
H_{LL}&=&N\hbar\omega_\perp - \frac{\hbar^2}{2m}\sum_{i=1}^N\frac{\partial^2}{\partial z_i^2}
+g_{1D}\sum_{i<j}\delta(z_i-z_j)
\nonumber \\
&+&\sum_{i=1}^N \frac{m\omega_z^2 z_i^2}{2} \;.
\label{hamLL}
\end{eqnarray}
When the number of particles is large, the properties of the ground state of the Hamiltonian (\ref{hamLL}) coincide 
with the ones obtained from the LL equation of state within LDA. However, for small systems one expects deviations 
and the DMC method provides us with a powerful tool. The relevant parameters are the same as for the 3D simulation 
with the Hamiltonian (\ref{ham}): the number of particles $N$, the ratio $a/a_\perp$ fixing the strength of the 
contact potential $g_{1D}$ and the anisotropy parameter $\lambda$ fixing the strength of the longitudinal confinement 
in units of $\hbar\omega_\perp$. Importance sampling is realized through the trial wavefunction 
$\psi_T(z_1,..,z_N)=\prod_i g_1(z_i)\prod_{i<j}g_2(z_i-z_j)$.  The one-body term is of gaussian form
$g_1(z)=\exp(-\alpha_z z^2)$ and the two-body term has been chosen as $g_2(z)=\cos[k(|z|-L)]$ for $|z|\le L$ and
$g_2(z)=1$ for $|z|>L$. The condition $k\tan(kL)=a/a_\perp^2$ fixes the proper boundary conditions of the two-body 
problem in $z=0$ related to the $\delta$-potential. The parameters $\alpha_z$ and $L$ are variational parameters.
We notice that our choice of the two-body Jastrow factor reproduces both the weakly interacting and the Tonks regime.
In fact, if $kL$ is small, $g_2(z=0)\simeq 1$ and the contact potential is almost transparent.  On the other hand, if 
$kL$ approaches $\pi/2$, $g_2(z=0)\simeq 0$ and the contact potential behaves as an impenetrable barrier.  
 
\section{Results}

We first consider a system of very few particles ($N=5$) and we
consider different values of the ratio $a/a_\perp$. Figs. 1-2 refer to $a/a_\perp=0.2$, and we present results for the
energy per particle and the mean square radius of the cloud in the longitudinal direction as a function of the anisotropy
parameter $\lambda=\omega_z/\omega_\perp$ . Results from the GP equation (\ref{GP}) and from the Lieb-Liniger equation of 
state in LDA are also shown. We find that the HS and SS potential give practically the same results even 
for the largest values of $\lambda$, showing that for these parameters we are well within the universal regime where the
details of the potential are irrelevant. For large values of $\lambda$ the DMC results agree well with the predictions of
GP equation. By decreasing $\lambda$ beyond mean-field effects become visible and both the energy per particle and the 
mean square radius approach the LL result when $N\lambda a_\perp^2/a^2\sim 1$, corresponding to $\lambda\sim 10^{-2}$. 
Finally, for the smallest values of the anisotropy parameter ($\lambda\sim 10^{-4}$) we find clear evidence of the Tonks 
gas behavior both in the energy and in the shape of the cloud. It is worth stressing that beyond mean-field effects 
occurring in the small $\lambda$ regime can be only obtained by using DMC. A Variational Monte-Carlo (VMC) calculation 
based on the trial wavefunction $\psi_T({\bf R})$ described above, would yield results in good agreement with mean-field 
over the whole range of values of $\lambda$. DMC results using the Lieb-Liniger Hamiltonian $H_{LL}$ of Eq. (\ref{hamLL}) 
are also shown and coincide with the results of the 3D Hamiltonian (\ref{ham}). This shows that the 3D 
interatomic potential is correctly described by the 1D $\delta$-potential even for the largest values of $\lambda$. In 
fact, due to the small number of particles, the density profile of the cloud in the transverse direction is correctly 
described by the harmonic oscillator ground-state wavefunction (see Fig. 9). The 1D character of the system is also 
evident from Fig. 1 which shows that $E/N-\hbar\omega_\perp$ is always smaller than the transverse confining energy. 
Deviations of DMC results from the LL equation of state arise because of finite size effects. These effects become 
less and less important as $\lambda$ decreases and one enters the regime $(E/N-\hbar\omega_\perp)/
\hbar\omega_\perp \gg\lambda$ where LDA applies. In terms of the mean square radius of the cloud (see Fig. 2),
the condition of applicability of LDA requires $\langle z^2\rangle^{1/2}$ much larger than the corresponding 
ideal gas (IG) value.  

\begin{figure}
\begin{center}
\includegraphics*[width=0.95\columnwidth,height=0.6\columnwidth]{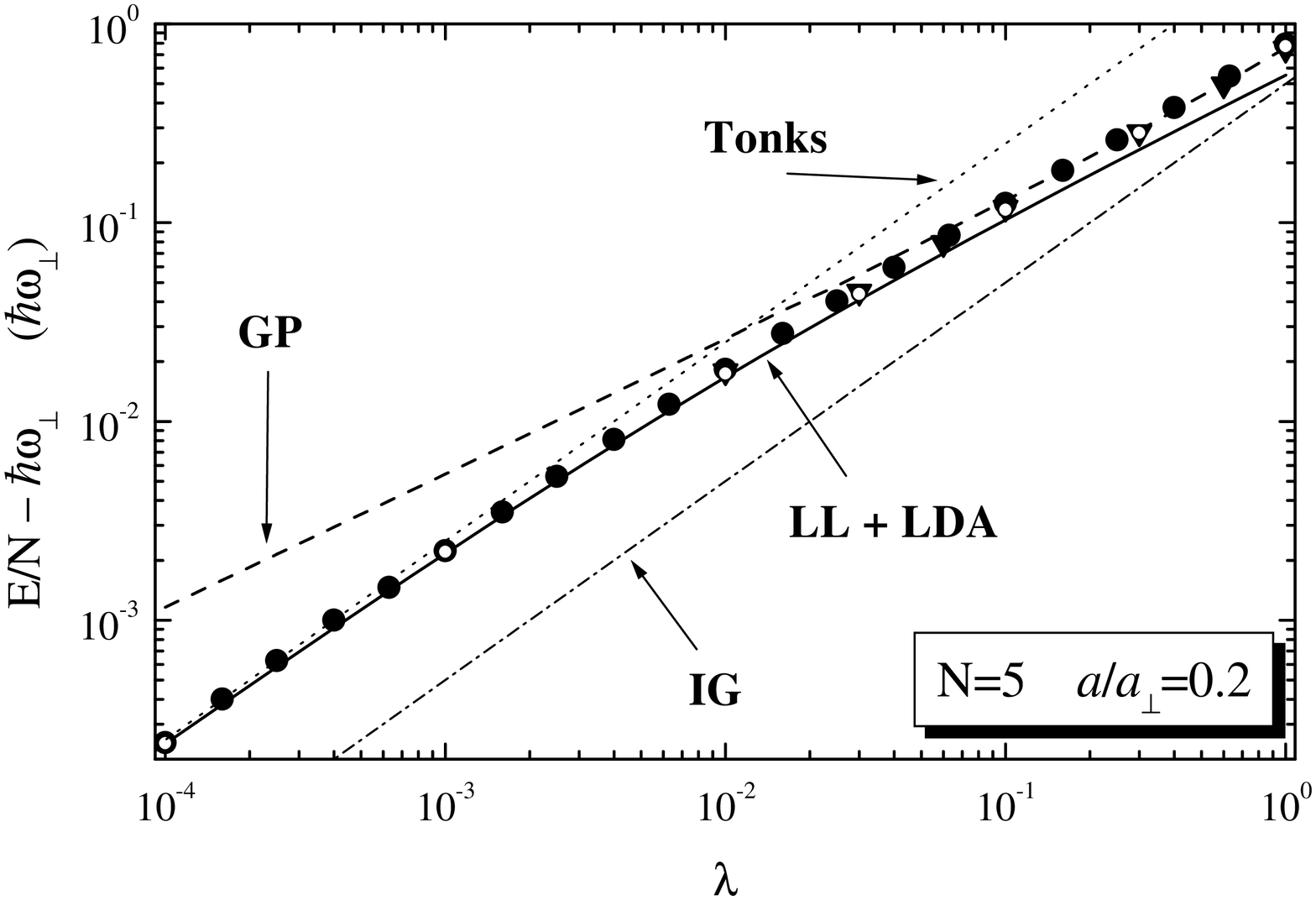}
\vspace{3 mm}
\caption{Energy per particle as a function of $\lambda$. DMC results: HS potential (solid circles), SS potential (solid 
triangles), LL Hamiltonian (\ref{hamLL}) (open circles). Dashed line: GP equation (\ref{GP}), solid line: LL equation 
of states in LDA, dotted line: gas of Tonks, dot-dashed line: non-interacting gas. Error bars are smaller than the size 
of the symbols.}
\label{Fig1}
\end{center}
\end{figure}

In Figs. 3-4 we present results for $a/a_\perp=0.04$, corresponding to a less tight transverse confinement or, equivalently, 
to a smaller scattering length. By decreasing $a/a_\perp$ we enter more deeply in the universal regime where the theory of 
pseudo-potentials applies and we find no difference between HS and SS results. 
Further, as for the $a/a_\perp=0.2$ case, we find no difference between DMC results obtained starting from Hamiltonian 
(\ref{ham}) and from the 1D Hamiltonian (\ref{hamLL}). Compared to Figs. 1-2, the cross-over between mean field and 
Lieb-Liniger occurs for smaller values of $\lambda$ . For the smallest values of $\lambda$ beyond mean-field effects 
become evident, though one would need to decrease $\lambda$ even further to enter the Tonks gas regime.      

\begin{figure}
\begin{center}
\includegraphics*[width=0.95\columnwidth,height=0.6\columnwidth]{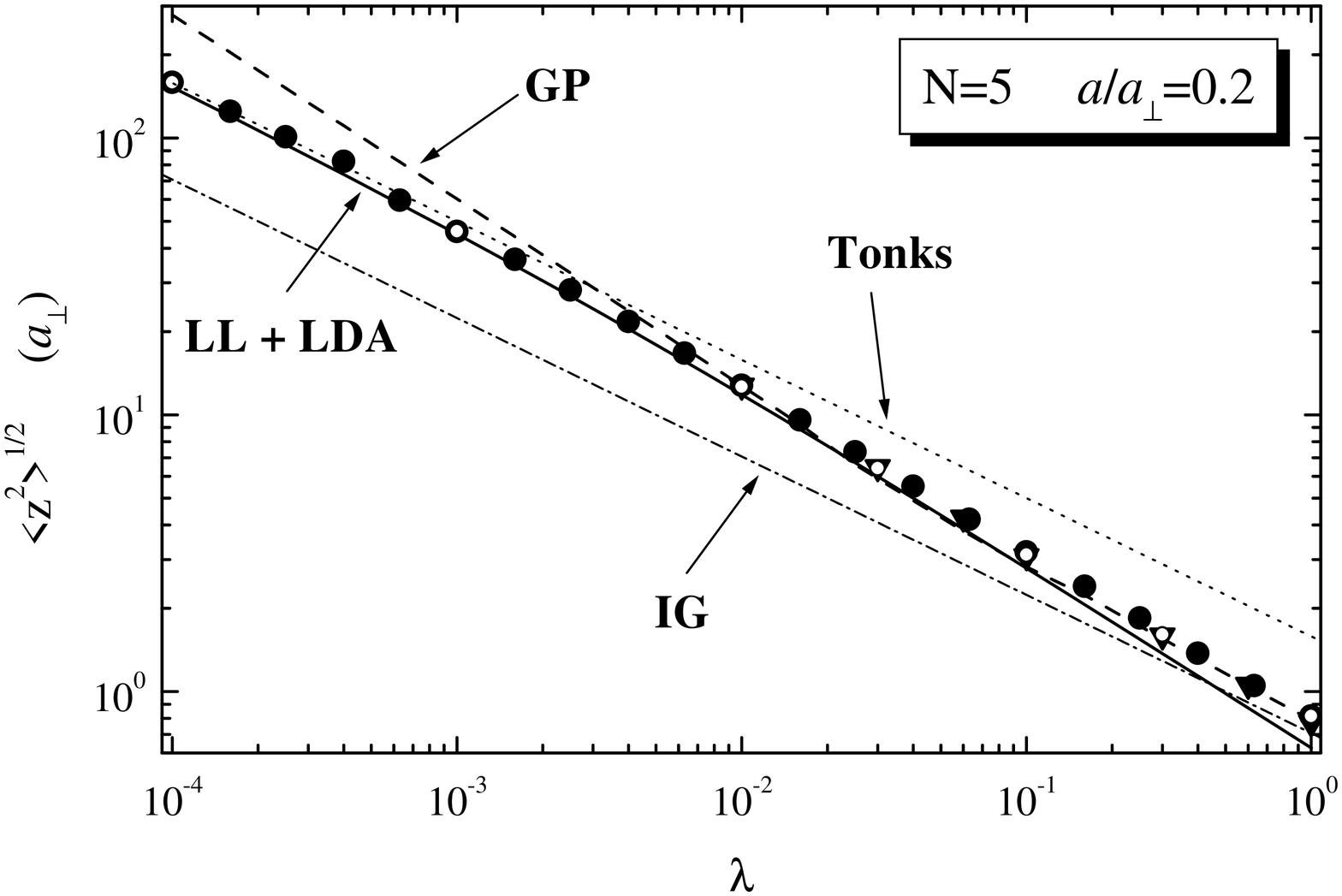}
\vspace{3 mm}
\caption{Mean square radius along $z$ as a function of $\lambda$. Symbols and line codes are the same as in Fig. 1. 
Error bars are smaller than the size of the symbols.}
\label{Fig2}
\end{center}
\end{figure}

\begin{figure}
\begin{center}
\includegraphics*[width=0.95\columnwidth,height=0.6\columnwidth]{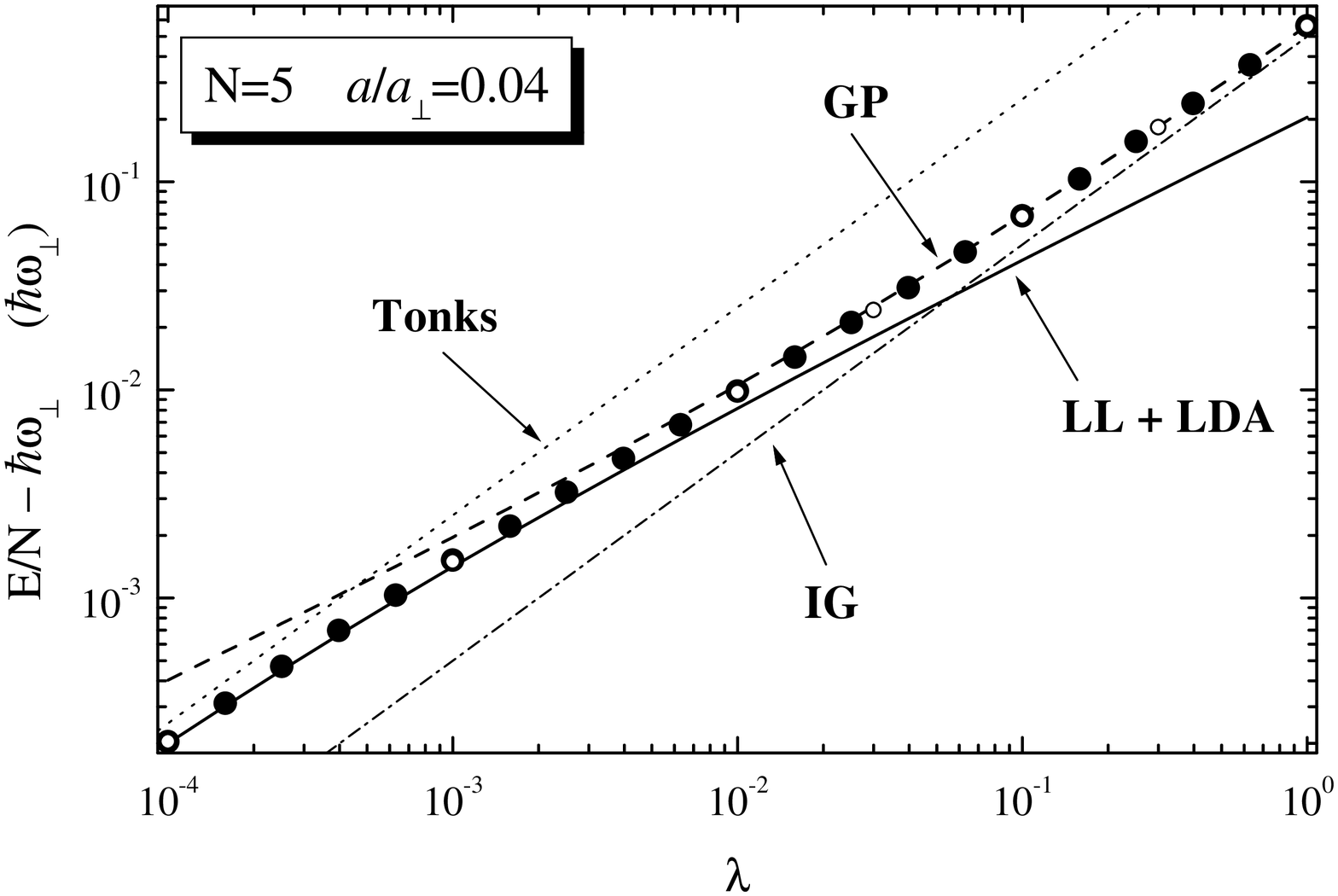}
\vspace{3 mm}
\caption{Energy per particle as a function of $\lambda$. DMC results: HS potential (solid circles), LL Hamiltonian 
(\ref{hamLL}) (open circles). Line codes are the same as in Fig. 1.}
\label{Fig3}
\end{center}
\end{figure}

\begin{figure}
\begin{center}
\includegraphics*[width=0.95\columnwidth,height=0.6\columnwidth]{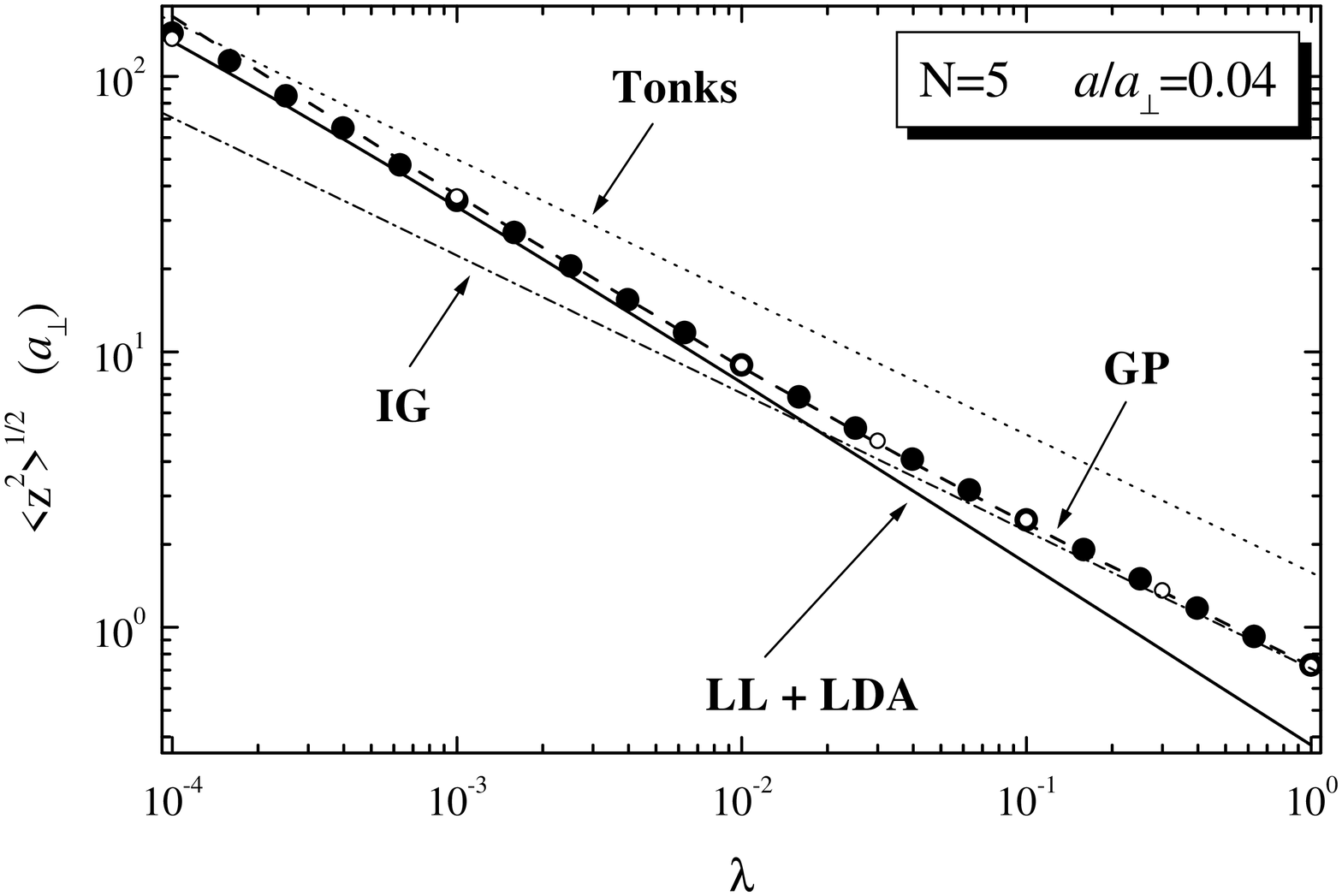}
\vspace{3 mm}
\caption{Mean square radius along $z$ as a function of $\lambda$. Symbols and line codes are the same as in Fig. 3.}
\label{Fig4}
\end{center}
\end{figure}

The results for $a/a_\perp=1$ are shown in Figs. 5-6. In this case the HS and SS potential give significantly different 
results in the large $\lambda$ regime. For both potentials the mean-field description is inadequate. By decreasing $\lambda$
the HS system enters the Tonks regime before approaching the LL results, whereas the SS system crosses from the ideal gas 
(IG) regime to the Tonks regime.  For this value of $a/a_\perp$ the DMC results with the Hamiltonian (\ref{hamLL}) coincide
exactly with the ones of the LL equation of state in LDA due to the large coupling constant.

\begin{figure}
\begin{center}
\includegraphics*[width=0.95\columnwidth,height=0.6\columnwidth]{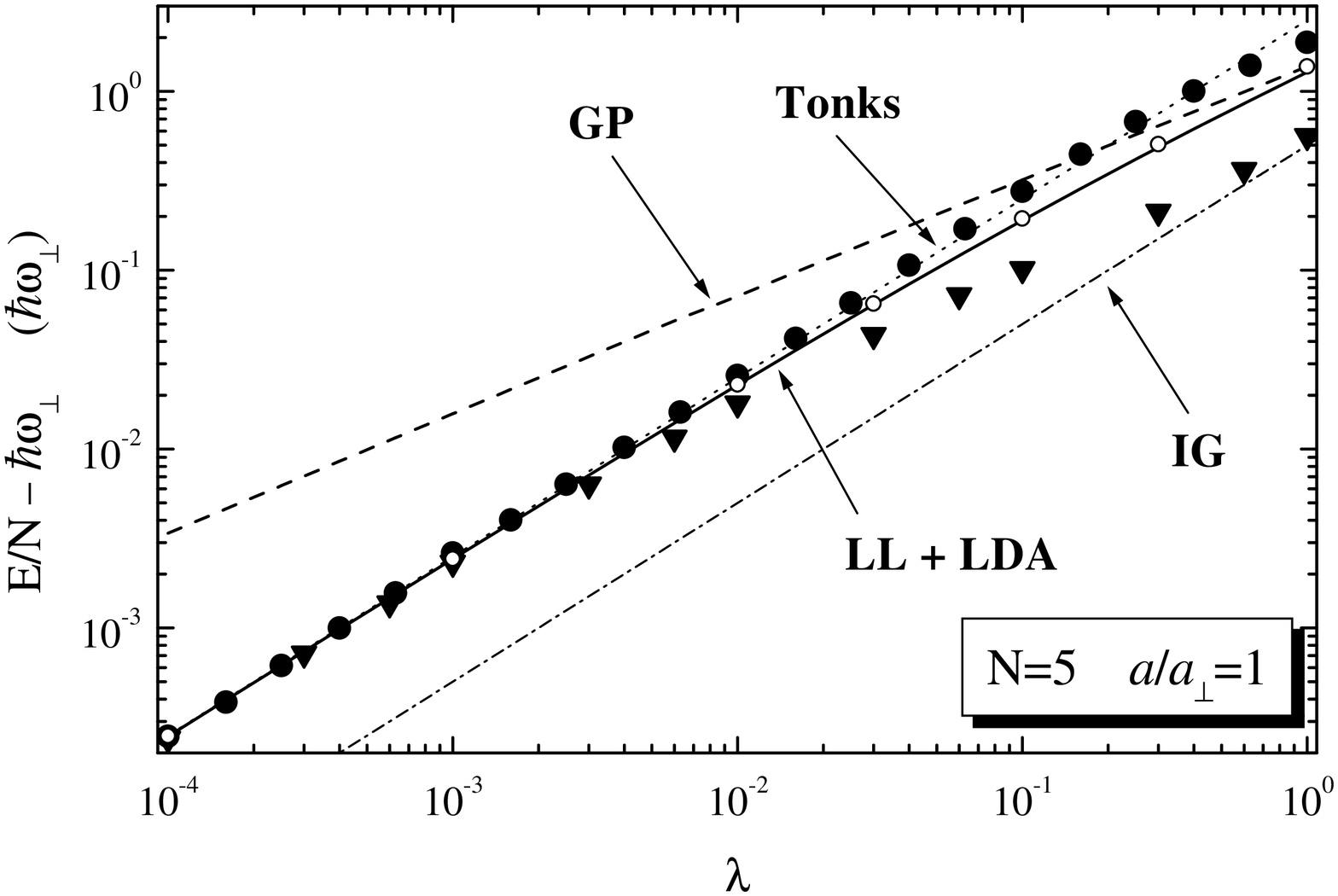}
\vspace{3 mm}
\caption{Energy per particle as a function of $\lambda$. Symbols and line codes are the same as in Fig. 1.}
\label{Fig5}
\end{center}
\end{figure}

\begin{figure}
\begin{center}
\includegraphics*[width=0.95\columnwidth,height=0.6\columnwidth]{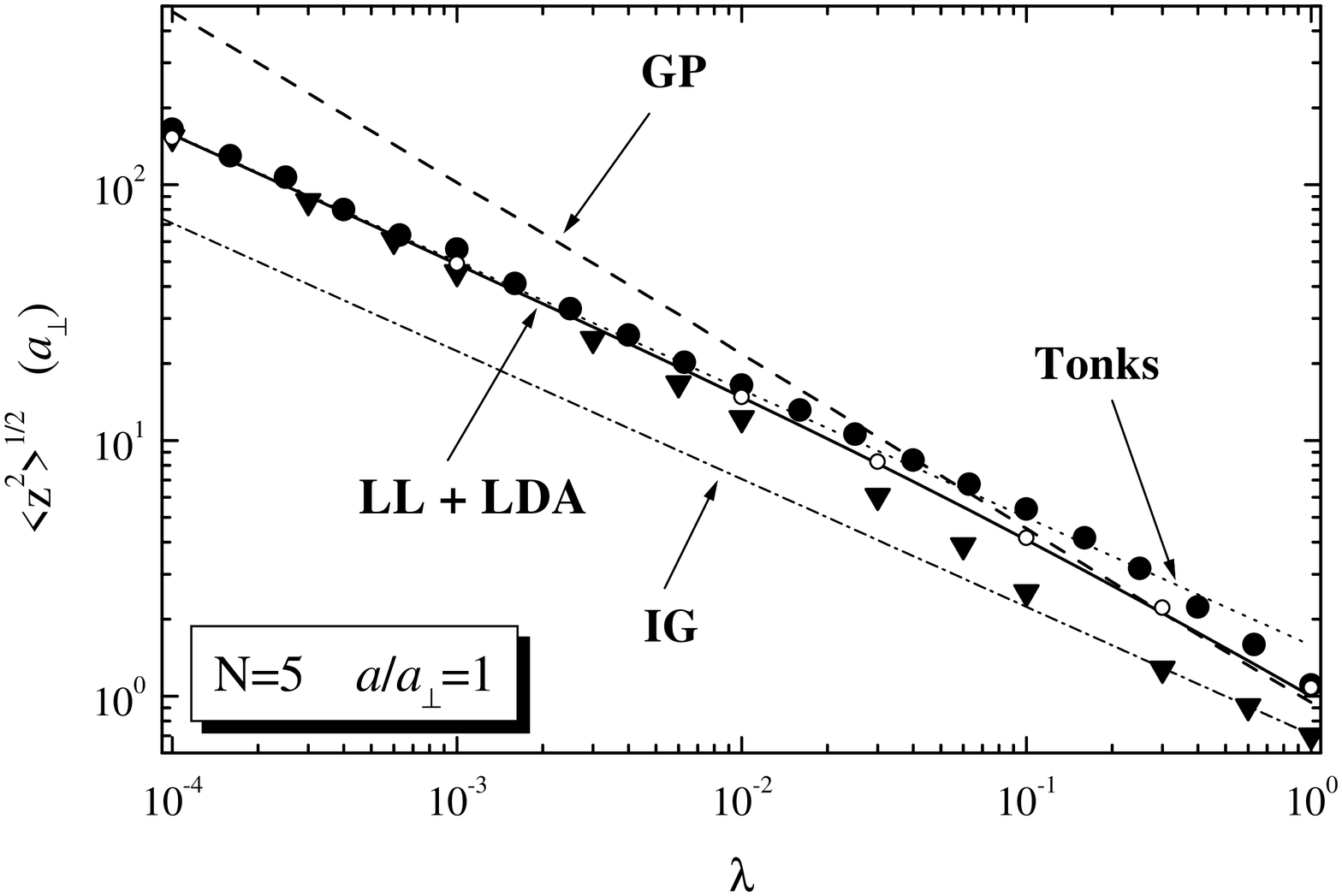}
\vspace{3 mm}
\caption{Mean square radius along $z$ as a function of $\lambda$. Symbols and line codes are the same as in Fig. 1.}
\label{Fig6}
\end{center}
\end{figure}

Figs. 7-8 refer to a much larger system with $N=100$ and $a/a_\perp=0.2$. In this case, we see a clear cross-over
from 3D mean field, at large $\lambda$, to 1D LL at small $\lambda$. Important beyond mean-field effects become
evident in the energy per particle as $N\lambda a_\perp^2/a^2\sim 1$, corresponding to $\lambda\sim 10^{-3}$.
The Tonks regime would correspond to even smaller values of $\lambda$ which are difficult to obtain in our 
simulation. However, for the smallest values of $\lambda$ reported in Fig. 7 we find already very good 
agreement with the LL equation of state. One should notice that small deviations from mean field are also visible 
for $\lambda\sim 1$, and are due to high density corrections to the GP equation. The DMC results with the 
1D Hamiltonian (\ref{hamLL}) follow exactly the LDA prediction showing that 
the deviations seen in Figs. 1-2 are due to finite size effects. In the cross-over region from the mean-field
to the 1D LL regime, residual 3D effects are still present (see Fig. 9) and produce small deviations from the
LL equation of state. 

\begin{figure}
\begin{center}
\includegraphics*[width=0.95\columnwidth,height=0.6\columnwidth]{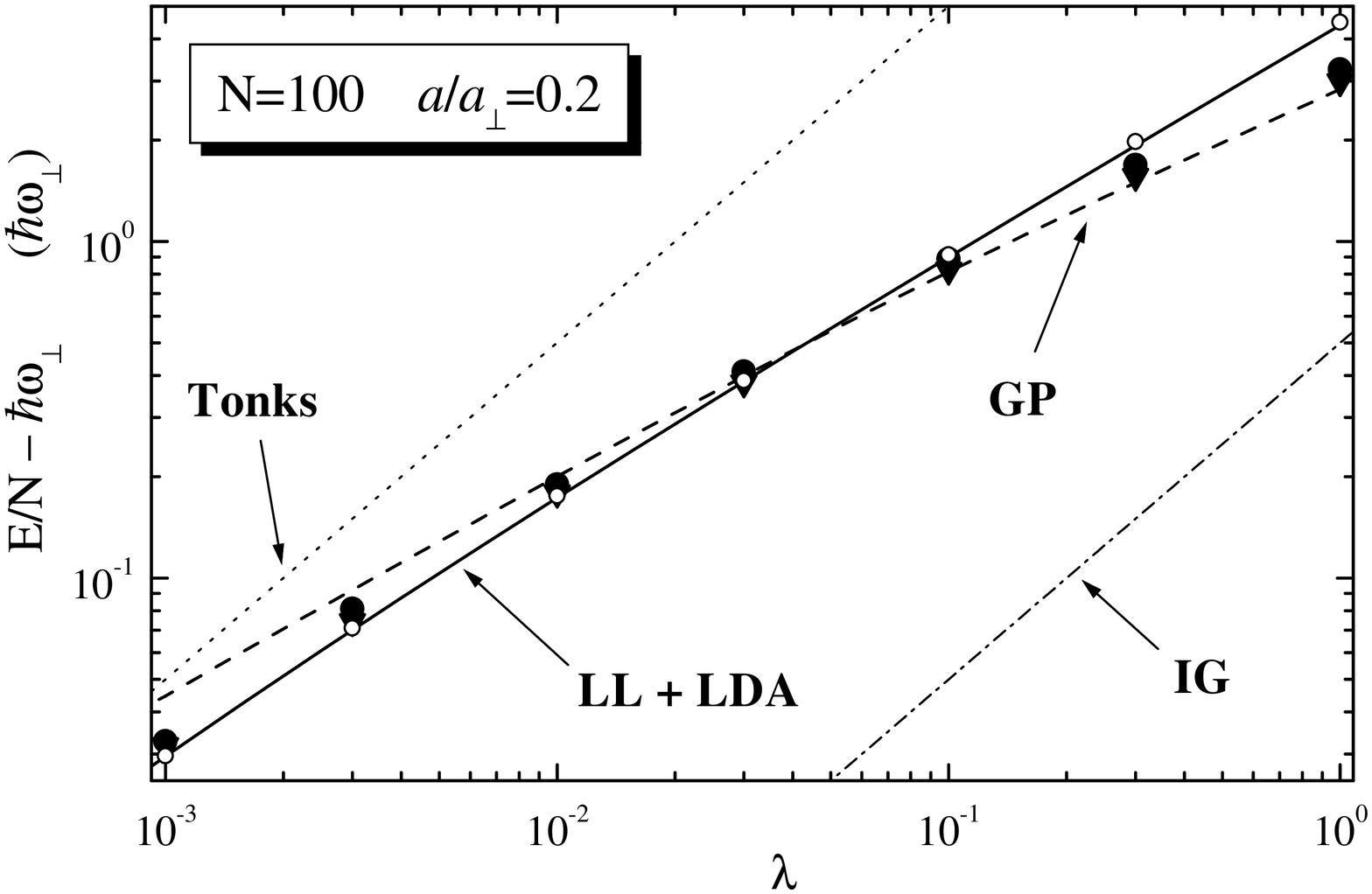}
\vspace{3 mm}
\caption{Energy per particle as a function of $\lambda$. Symbols and line codes are the same as in Fig. 1.}
\label{Fig7}
\end{center}
\end{figure}

\begin{figure}
\begin{center}
\includegraphics*[width=0.95\columnwidth,height=0.6\columnwidth]{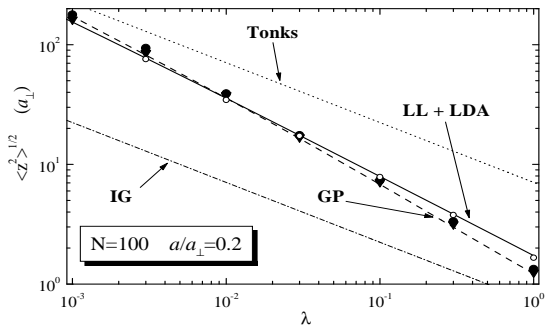}
\vspace{3 mm}
\caption{Mean square radius along $z$ as a function of $\lambda$. Symbols and line codes are the same as in Fig. 1.}
\label{Fig8}
\end{center}
\end{figure}

Finally, in Fig. 9, we show results for the mean square radius in the transverse direction. The cross-over from 
3D to 1D is clearly visible in the case of $N=100$, for both the HS and SS potential, and for the HS potential 
in the case of $N=5$ and $a/a_\perp=1$. For the system with $N=5$ and $a/a_\perp=0.2$ we only see small deviations 
from $\sqrt{\langle r_\perp^2\rangle}=a_\perp$ for the largest values of $\lambda$. In the $a/a_\perp=0.04$, as 
well as in the $a/a_\perp=1$ case with the SS potential, the transverse density profile is well described by the
harmonic oscillator wavefunction and we find $\sqrt{\langle r_\perp^2\rangle}\simeq a_\perp$ over the whole range
of values of $\lambda$. It is worth noticing that for the $N=5$ system with SS potential, the largest deviations
from $\sqrt{\langle r_\perp^2\rangle}=a_\perp$ are achieved for $a/a_\perp=0.2$, corresponding to a transverse 
confinement $a_\perp=R$ where $R$ is the range of the SS potential.

\begin{figure}
\begin{center}
\includegraphics*[width=0.95\columnwidth,height=0.6\columnwidth]{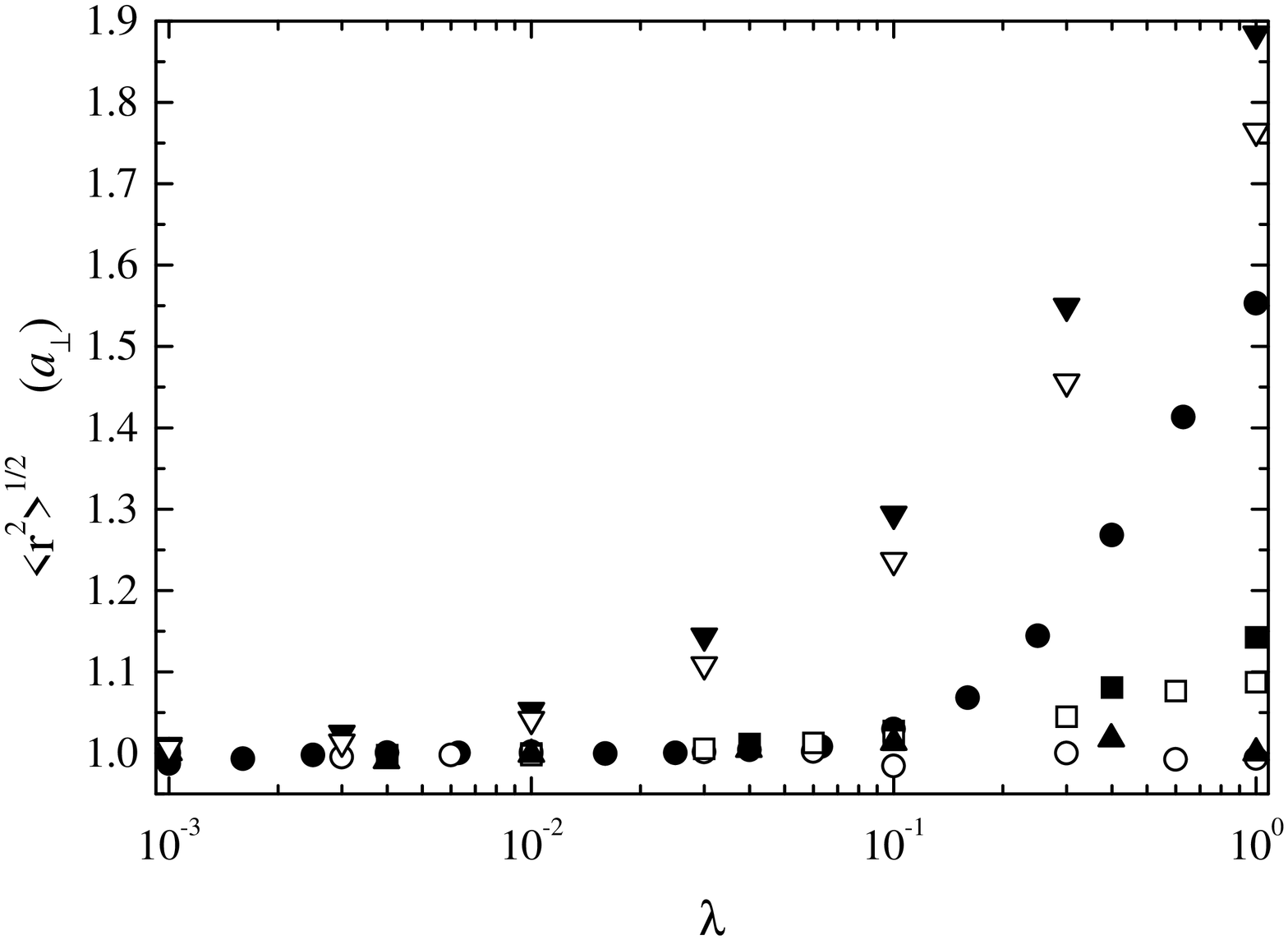}
\vspace{3 mm}
\caption{Mean square radius in the radial direction as a function of $\lambda$. Solid symbols: HS potential; 
open symbols: SS potential. Down triangles: $N=100$ and $a/a_\perp=0.2$; circles: $N=5$ and $a/a_\perp=1$;
squares: $N=5$ and $a/a_\perp=0.2$; up triangles: $N=5$ and $a/a_\perp=0.04$. Error bars are smaller than the 
size of the symbols.}
\label{Fig9}
\end{center}
\end{figure}

\section{Conclusions}
In this paper we present exact Quantum Monte Carlo results of the ground-state energy and structure of a Bose 
gas confined in highly anisotropic harmonic traps. Starting from a 3D Hamiltonian, where interparticle 
interactions are model by a hard-sphere or a soft-sphere potential, we show that the system exhibits striking
features due to particle correlations. By reducing the anisotropy parameter $\lambda$, while the number of 
particles $N$ and the ratio $a/a_\perp$ of scattering to transverse oscillator length are kept fixed, the 
system crosses from a regime where mean-field theory applies to a regime which is well described by the 1D
Lieb-Liniger equation of state in local density approximation. In the cross-over region both theories fail 
and one must resort to exact methods to account properly for both finite size effects and residual 3D effects. 
For very small values of $\lambda$ we find clear evidence, both in the energy per particle and in the 
longitudinal size of the cloud, of the fermionization of the system in the Tonks-Girardeau regime.   
  
\section{ACKNOWLEDGMENTS}
We gratefully acknowledge valuable discussions with J. Boronat, C. Menotti and S. Stringari. This research is 
supported by Ministero dell'Istruzione, dell'Universit\`a e della Ricerca (MIUR).

{\it Note added.} While this work was being prepared for publication, a preprint by D. Blume \cite{Blume} 
appeared in which the author also calculates the ground-state properties of Bose gases in highly anisotropic 
traps using Diffusion Monte Carlo techniques. The results obtained are similar to ours.


\begin{references}

\bibitem{Schwartz} M. Schwartz, Phys. Rev. B {\bf 15}, 1399 (1977); F.D.M. Haldane, Phys. Rev. Lett. {\bf 47}, 1840 
(1981).

\bibitem{Girardeau} M. Girardeau, J. Math. Phys. (N.Y.) {\bf 1}, 516 (1960).

\bibitem{Mit} A. G\"orlitz {\it et al.}, Phys. Rev. Lett. {\bf 87}, 130402 (2001).

\bibitem{Paris} F. Schreck {\it et al.}, Phys. Rev. Lett. {\bf 87}, 080403 (2001).

\bibitem{Konstanz} S. Dettmer {\it et al.}, Phys. Rev. Lett. {\bf 87}, 160406 (2001).

\bibitem{Olshanii} M. Olshanii, Phys. Rev. Lett. {\bf 81}, 938 (1998).

\bibitem{Petrov} D.S. Petrov, G.V. Shlyapnikov and J.T.M. Walraven, Phys. Rev. Lett. {\bf 85}, 3745 (2000).

\bibitem{Lieb} E.H. Lieb and W. Liniger, Phys. Rev. {\bf 130}, 1605 (1963); E.H. Lieb, Phys. Rev. {\bf 130}, 1616
(1963).

\bibitem{Girardeau2} M.D. Girardeau, E.M. Wright and J.M. Triscari, Phys. Rev. A {\bf 63}, 033601 (2001).

\bibitem{Dunjko} V. Dunjko, V. Lorent and M. Olshanii, Phys. Rev. Lett. {\bf 86}, 5413 (2001).

\bibitem{Menotti} C. Menotti and S. Stringari, preprint cond-mat/0201158.

\bibitem{Girardeau3} K.K. Das, M.D. Girardeau and E.M. Wright, preprint cond-mat/0204317.

\bibitem{US} S. Giorgini, J. Boronat and J. Casulleras, Phys. Rev. A {\bf 60}, 5129 (1999).

\bibitem{BORO} For a general reference of the DMC method see for example J. Boronat and J. Casulleras, Phys. Rev. B
{\bf 49}, 8920 (1994). Mean square radii are calculated using the pure estimator technique developed in 
J. Casulleras and J. Boronat, Phys. Rev. B {\bf 52}, 3654 (1995).

\bibitem{Esry} B.D. Esry, Phys. Rev. A {\bf 55}, 1147 (1997).

\bibitem{Blume} D. Blume, preprint cond-mat/0206244.

\end{references}
\end{document}